\documentclass[journal]{IEEEtran}

\usepackage{amssymb}
\usepackage{amsmath}
\usepackage{amsfonts}
\usepackage{dsfont} 
\usepackage[utf8]{inputenc}
\usepackage{amsthm}
\usepackage{cite}
\usepackage{subfigure}
\usepackage{color}
\usepackage{times}
\usepackage{graphicx}
\usepackage{psfrag}
\usepackage[dvipsnames]{xcolor}

\makeatletter
\newcommand*\rel@kern[1]{\kern#1\dimexpr\macc@kerna}
\newcommand*\widebar[1]{%
  \begingroup
  \def\mathaccent##1##2{%
    \rel@kern{0.8}%
    \overline{\rel@kern{-0.8}\macc@nucleus\rel@kern{0.2}}%
    \rel@kern{-0.2}%
  }%
  \macc@depth\@ne
  \let\math@bgroup\@empty \let\math@egroup\macc@set@skewchar
  \mathsurround\z@ \frozen@everymath{\mathgroup\macc@group\relax}%
  \macc@set@skewchar\relax
  \let\mathaccentV\macc@nested@a
  \macc@nested@a\relax111{#1}%
  \endgroup
}
\makeatother

\newcommand{\bm}{\mathbf}
\newcommand{\be}{\begin{equation}}
\newcommand{\ee}{\end{equation}}
\newcommand{\bea}{\begin{eqnarray}}
\newcommand{\eea}{\end{eqnarray}}

\newcommand{\z}{{\bm z}}
\newcommand{\p}{{\bm p}}

\newcommand{\ba}{{\bm a}}

\newcommand{\bA}{{\bm A}}
\newcommand{\bI}{{\bm I}}

\newcommand{\bW}{{\bm W}}

\newcommand{\bG}{{\bf G}}
\newcommand{\bD}{{\bf D}}
\newcommand{\bC}{{\bf C}}

\newcommand{\bH}{{\bf H}}

\newcommand{\bh}{{\bf h}}
\newcommand{\bz}{{\bf z}}

\newcommand{\bs}{{\bf s}}

\IEEEoverridecommandlockouts
\begin{document}

\title{FBMC Receiver Design and Analysis for Medium and Large Scale Antenna Systems}

\author{\normalsize Hamed Hosseiny$^\dagger$, Arman Farhang$^*$ and Behrouz Farhang-Boroujeny$^\dagger$  
\\$^\dagger$ECE Department, University of Utah, USA, \\
$^*$Department of Electronic Engineering, Maynooth University, Ireland. \\
Email: \{hamed.hosseiny, farhang\}@utah.edu, \{arman.farhang\}@mu.ie}
\maketitle

\begin{abstract}

In this paper, we design receivers for filter bank multicarrier-based (FBMC-based) massive MIMO considering practical aspects such as channel estimation and equalization. In particular, we propose a spectrally efficient pilot structure and a channel estimation technique in the uplink to jointly estimate all the users' channel impulse responses. We mathematically analyze our proposed channel estimator and find the statistics of the channel estimation errors. These statistics are incorporated into our proposed equalizers to deal with the imperfect channel state information (CSI) effect. We revisit the channel equalization problem for FBMC-based massive MIMO, address the shortcomings of the existing equalizers in the literature, and make them more applicable to practical scenarios. The proposed receiver in this paper consists of two stages. In the first stage, a linear combining of the received signals at the base station (BS) antennas provides a coarse channel equalization and removes any multiuser interference. In the second stage, a per subcarrier fractionally spaced equalizer (FSE) takes care of any residual distortion of the channel for the user of interest. We propose an FSE design based on the equivalent channel at the linear combiner output. This enables the applicability of our proposed technique to small and/or distributed antenna setups such as cell-free massive MIMO. Finally, the efficacy of the proposed techniques is corroborated through numerical analysis.

\end{abstract}

\section{Introduction}
\IEEEPARstart{T}{he} emergence of new applications and technologies such as high data rate holographic communications and low latency high mobility communications for autonomous driving, as well as massive machine-type communications, has marked a new era in communications, \cite{tataria20206g}. This calls for the development of a flexible air interface in different levels and the associated modulation scheme to deliver unprecedented levels of connectivity, reliability, and flexibility, often supplemented with strict latency requirements. Even though orthogonal frequency division multiplexing (OFDM) has been chosen for the physical layer of both 4G and 5G systems, its shortcomings, such as its sensitivity to frequency errors and bandwidth efficiency loss due to the redundant cyclic prefix (CP) and high out-of-band emissions motivate consideration of other alternative waveforms. Its high spectral efficiency and flexibility in prototype filter design to serve a diverse set of applications in future networks make filter bank multicarrier (FBMC) a promising candidate waveform for future wireless systems \cite{farhang2011ofdm,nissel2017filter}.

Design of novel physical layer technologies and flexible modulation schemes such as FBMC that are bolstered with advanced co-located/distributed multiple antenna systems contribute towards an increased capacity and reliability in the future multiuser networks. The authors in \cite{farhang2014filter} and \cite{aminjavaheri2015frequency} started this line of research and demonstrated the self-equalization/channel-flattening effect of FBMC in massive MIMO channels. In a more recent work, \cite{aminjavaheri2018filter}, it was noted that the channel-flattening effect of FBMC is limited and, thus, a more accurate equalization method was needed.  In particular, it was shown that this limit is the result of the correlation between combiner taps and channel impulse responses between the user terminal and the base station (BS) antennas, \cite{aminjavaheri2018filter}. It was further shown that this correlation converges to an equivalent channel which resembles the power delay profile (PDP) of the set of channels between the user of interest and the BS antennas.  This finding was then used to propose a per-subcarrier per-user equalizer that recovers the Nyquist property that was broken by this PDP channel. 

The authors in \cite{rottenberg2018performance} use concepts in random matrix theory to obtain the asymptotic performance of FBMC-based massive MIMO systems with a linear combiner, similar to those in \cite{farhang2014filter} and \cite{aminjavaheri2015frequency}, and presented the mean squared error (MSE) of the recovered data symbols. These results show that the MSE saturates to a lower bound and  becomes uniform across all the subcarriers. Taking note that the MSE and signal-to-interference plus noise (SINR) are inversely related, one may realize that this result is inline with the results of \cite{aminjavaheri2018filter}.
 The authors in \cite{singh2019uplink}, on the other hand, analyze the performance of FBMC in multiuser massive MIMO systems with co-located antennas and derive lower bound expressions for achievable sum-rates with and without perfect channel state information (CSI) in the uplink when linear combiners are deployed. The results in this paper are also in line with those in  \cite{aminjavaheri2018filter}, and \cite{rottenberg2018performance}.

It is well known that, accurate estimates of the underlying channels are required at the BS to deliver the promising benefits of massive MIMO systems. Furthermore, many of the emerging applications in future wireless networks require ultra-reliable low-latency communications (URLLC), \cite{schulz2017latency}. This necessitates the need for highly accurate channel estimation techniques with minimal training overheads, especially in multiuser scenarios. In addition, the current literature on FBMC-based massive MIMO is mainly focused on asymptotic analysis, without much focus on the practical aspects, \cite{aminjavaheri2018filter,singh2019uplink,rottenberg2018performance}. In particular, to the best of our knowledge, there is no published work to date that extensively addresses the practical problems of channel estimation that were noted above, e.g., equalization in the presence of imperfect CSI, as well as the scenarios where the number of BS antennas are limited.

Hence, the focus of this paper is on the design of a practical receiver for FBMC-based massive MIMO systems in the presence of imperfect CSI. We start with the channel estimation problem in the uplink of FBMC-based network. We note that, in FBMC, subcarriers are orthogonal in the real field. This makes channel estimation a more complex task when compared to its OFDM counterpart. Existing channel estimation methods for FBMC are mainly based on the interference approximation method (IAM), e.g., see \cite{lele2008channel}. IAM is a frequency domain channel estimation technique and requires the maximum channel delay spread to be much shorter than the symbol interval. Thus, when this condition does not hold, IAM leads to inaccurate channel estimates. To avoid this issue, the time domain channel estimation techniques were proposed \cite{kong2014time,caus2012transmitter,kofidis2015preamble,singh2019time}. The authors in \cite{kong2014time} and \cite{caus2012transmitter} propose time domain channel estimation techniques where guard symbols are required to separate different users' pilots. This leads to a spectral efficiency loss and limitations in terms of latency, especially as the number of users scales up. Hence, in multiuser scenarios, reducing the pilot overhead for channel estimation in FBMC systems is of a paramount importance. An alternative time domain channel estimation method for FBMC and its extension to MIMO channels was proposed in \cite{kofidis2015preamble} and further studied in \cite{singh2019time}. This method considers sending pilots for each user on all the subcarriers where the users' pilots are multiplexed in the code domain to allow sharing the same time-frequency resources for channel estimation. However, this solution suffers from a large amount of computational load at the receiver for demultiplexing different users' channel responses based on their code sequences.

To address the aforementioned issues for channel estimation and meet the stringent latency requirements of future networks, in this paper, we propose a pilot structure and a time domain channel estimation method for FBMC-based massive MIMO. Opposed to the existing literature, our proposed pilot structure interleaves different users' pilots in time and frequency without any guard symbols between them. We show that for a given user with the channel length $L$, utilization of only $L$ pilots for channel estimation is sufficient. This clearly leads to a great amount of savings in signaling overhead. This, in turn, translates into a reduced latency and improved spectral efficiency, especially as the number of users increases. Furthermore, this brings a significant relaxation on the pilot contamination problem in massive MIMO networks, \cite{rusek2012scaling}, given that the minimum number of pilots is assigned to each user. The proposed channel estimator takes advantage of the intrinsic interference due to the absence of guard symbols between different users' pilots and jointly estimates all the users' channel impulse responses. We also mathematically analyze our proposed channel estimator and obtain the statistics of the channel estimation errors, which prove to be useful in combatting the imperfect CSI effect. 
%

We also address the channel equalization problem in massive MIMO FBMC, more inclined towards practical scenarios with both co-located and distributed antenna setups, i.e., cell-free MIMO \cite{ngo2017cell}. We note that the PDP equalization method that has been proposed in \cite{aminjavaheri2018filter} may not be applicable to such cases. By calculating the equivalent channel that is seen at the linear combiner output, we propose a novel fractionally spaced equalizers (FSE) as the second equalization stage and show its excellent performance using simulations. The proposed FSE in this paper also serves as a benchmark for evaluating the effectiveness of the PDP-based equalizer in \cite{aminjavaheri2018filter}. Our analyses in this paper also include a study of the effects of the channel estimation errors on both the PDP-based and equivalent channel-based equalizers. Furthermore, we have developed methods for compensating the effect of imperfect CSI in massive MIMO setups with both co-located and distributed antennas. We note that the proposed PDP equalizer in \cite{aminjavaheri2018filter} assumes perfect knowledge of the underlying channel PDP at the receiver. Clearly, this is not the case in practice, and hence, an estimate of the channel PDP should be obtained. Such estimate is also derived, and its efficacy is proven through simulations.

To summarize, the main contributions of this paper are the following; (i) We propose a joint multiuser and spectrally efficient channel estimation technique for the uplink of FBMC-based networks that is applicable to massive MIMO with both co-located and distributed antennas; (ii) We revisit the channel equalization problem in FBMC-based massive MIMO and propose a practical two-stage equalization technique that is highly effective in massive MIMO with both co-located and distributed antenna setups; (iii) We derive the statistical characteristics of the estimation errors of our proposed channel estimator and incorporate them into our proposed equalizers to tackle the imperfect CSI effects.
The rest of the paper is organized as follows. Section \ref{sec:FBMC} presents FBMC principles, paving the way towards presenting our proposed pilot structure and multiuser massive MIMO channel estimation technique in Section~\ref{sec:estimation}. In Section~\ref{sec:MAssive_MIMO}, the principles of FBMC for massive MIMO are explained along with an asymptotic analysis that sheds light on the need for an extra stage of equalization after resolving the multiuser interference and a coarse equalization of the channel in the first stage. We show that this equalizer should be a fractionally spaced one  and should be designed separately for each user. Two design techniques for the equalizers are also proposed. Section \ref{sec:imperfect} expands our analysis to the imperfect CSI and discusses the corresponding equalizers. Section \ref{sec:simulation} provides numerical analysis, confirming the validity of our claims through simulations. Finally, the paper is concluded in Section \ref{sec:conclusion}.

\vspace{3mm}
\noindent\textit{Notations:} Matrices, vectors and scalar quantities are denoted by boldface uppercase, boldface lowercase and normal letters, respectively.  $A(m,l)$ represents the element in the $m^{\rm th}$
row and the $l^{\rm th}$ column of $\bA$ and $\bA ^{-1}$ signifies the inverse of $\bA$. $\bI_M$ is the identity matrix of size $M \times M$, and $\bD = {\rm diag} ({\ba})$
is a diagonal matrix with diagonal elements in the vector $\ba$. Superscripts $(\cdot)^{-1}$, $(\cdot)^{\rm T}$, $(\cdot)^{\rm H}$ and $(\cdot)^*$ indicate inverse, transpose, conjugate transpose, and conjugate operations, respectively. $\mathfrak{R}\{\cdot\}$, $\mathfrak{I} \{\cdot\}$, $\mathds{E}\{\cdot\}$, ($\downarrow M$), $\star$ and ${\rm tr}\{\cdot\} $ represent real value, imaginary value, expectation, $M$ fold decimation, linear convolution and matrix trace operators, respectively. 
Finally, $\delta_{ij}$ represents the Kronecker delta function.


\section{FBMC Principles}
\label{sec:FBMC}
We consider the discrete time baseband equivalent of the staggered multi-tone (SMT) system. This modulation scheme divides the transmission bandwidth into $M$ sub-carrier bands with the normalized bandwidth of $1/M$ each. The real-valued data symbols in SMT are placed on a regular time-frequency grid with the time and frequency spacings of $T/2$ and $1/T$, respectively. Thus, the synthesized signal is expressed as
\begin{equation}
x[l] = \sum_{m=0}^{M-1} \sum_{n=-\infty}^{ \infty} s_{m,n} f_{m,n}[l] ,
\label{bbeq}
\end{equation}
where $s_{m,n}$ is the real-valued data symbol at the frequency index $m$ and the time index $n$, and
\begin{equation}\label{eq:fmn}
f_{m,n}[l] = f\big[l-n\frac{M}{2}\big] e^{j2\pi ml/M}e^{j\pi (m+n)/2},
\end{equation}
is the modulated and phase-adjusted pulse-shape that carries $s_{m,n}$. 

In \eqref{eq:fmn}, $f[l]$ is a prototype filter that is designed such that for all pairs of $(m,n)$ and $(m',n')$
\begin{equation}
    \Re \bigg\{ \sum_{l=- \infty}^{ \infty} f_{m,n}[l]  f^{*}_{m',n'}[l] \bigg\} = \delta_{mm'} \delta_{nn'}.
\end{equation}
This property, which is known as orthogonality in the real field, implies that  the set of functions $f_{m,n}[l]$, for all choices of $m$ and $n$, defines a basis set that carries the real-valued data symbols $s_{m,n}$. These data symbols can be extracted from the synthesized signal $x[l]$ by projecting $x[l]$ on the basis functions $f_{m,n}[l]$ and taking the real-part of the results. That is,
\begin{equation}
s_{m,n}=\Re\big\{ \langle x[l], f_{m,n}[l] \rangle\big\},
\end{equation}
and projection of $x[l]$ on $f_{m,n}[l]$ is defined as
\begin{equation}
\langle x[l], f_{m,n}[l] \rangle=\sum_{l=- \infty}^{ \infty} x[l]  f^{*}_{m,n}[l].
\end{equation}

Assuming a time-invariant channel, the received signal at the receiver can be written as
\begin{equation}
    r[l] = h[l] \star x[l] + \eta[l],
    \label{convolutionalform}
\end{equation}
where $h[l]$ represents multi-path channel impulse response with length $L$ and $\eta[l]$ is additive white Gaussian noise (AWGN) with the variance of $\sigma_{\eta}^2$, i.e., $\eta[l] \thicksim \mathcal{CN}(0,\sigma
_{\eta}^2)$. 

We define the analyzed/demodulated signal samples
\begin{equation}\label{eq:zmn}
z_{m,n} = \langle r[l], f_{m,n}[l] \rangle,
\end{equation}
and note that the data symbols $s_{m,n}$ are extracted by passing the sequence $z_{m,n}$ through an equalizer and taking the real-part of the output. In its simplest form, when the channel is approximated by a flat gain across each subcarrier band, a single-tap equalizer is sufficient. Since this approximation is not always valid,  a multi-tap equalizer may be favorable, \cite{hirosaki1980analysis}. For the equalizer to provide a satisfactory performance, accurate channel estimation is of a paramount importance. Thus, the following section is focused on channel estimation.

\section{ Channel Estimation} 
\label{sec:estimation}
\begin{figure}[t]
    \centering	\includegraphics[scale=0.25]{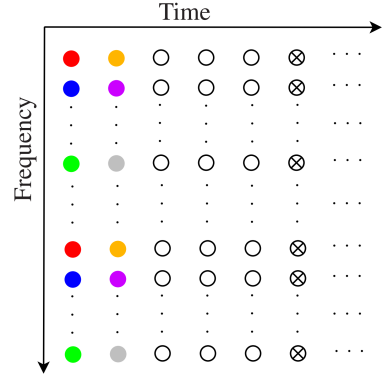}
		\caption{Proposed pilot structure for channel estimation in the uplink. Different users' pilots are specified with different colors. Crossed circles and empty circles represent data and zero symbols, respectively.}
		\label{pilot2user}
	\end{figure}

In this section, we propose a spectrally efficient pilot structure and a joint multiuser channel estimation technique for the uplink of FBMC-based networks. We also provide the estimation error characteristics of the proposed method. This will be used in the following sections to mitigate the imperfect CSI effects on the proposed channel equalization techniques. In our proposed channel estimation method, we use only $L$ rather than $M$ pilot subcarriers per user. Noting that typically $L\ll M$, using $L$ pilots significantly reduces the signaling overhead. Moreover, to further reduce the signaling overhead, guard bands are avoided between different users' pilot subcarriers. However, to avoid interference between the pilot and data symbols, we follow the previous literature, \cite{kong2014time}, and insert a few guard symbols after pilot symbols. The proposed pilot structure is presented in Fig.~\ref{pilot2user}.

As we perform per antenna channel estimation, for the sake of simplicity of the notations and without loss of generality, we drop the antenna index. We assume that $K$ users transmit their pilot symbols simultaneously. The pilot symbols that are transmitted by the $k$th user are put together in the real-valued column vector 
 $\p_k = [p_0^k,p_1^k, \ldots ,p^k_{N_{\rm p}-1}]^{\rm{T}}$, where $N_p$ is the number of pilots and the associated  demodulated vector at the receiver is  the complex-valued column vector $\bz_k = [z_{p^k_0},z_{p^k_1}, \ldots ,z_{p^k_{N_{\rm p}-1}}]^{\rm{T}}$.  

Following the SMT signal synthesis in \eqref{bbeq} and signal analysis in \eqref{eq:zmn}, straightforward manipulations leads to 
\begin{equation}
\bar{\bz} = \bar{\bA}\bar{\bh}+ \bar{\boldsymbol{\eta}},
\label{eq:linear_MU}
\end{equation}
where $\bar{\bz} = [\bz_0^{\rm T},\ldots,\bz_{K-1}^{\rm T}]^{\rm T}$, $\bar{\bh} = [\bh_0^{\rm T},\ldots,\bh_{K-1}^{\rm T}]^{\rm T}$, $\bar{\boldsymbol{\eta}} = [\boldsymbol{\eta}_0^{\rm T},\ldots,\boldsymbol{\eta}_{K-1}^{\rm T}]^{\rm T}$, $\bh_k=[h_k[0],\ldots,h_k[L-1]]^{\rm T}$ is the channel vector of user $k$, $\boldsymbol{\eta}_k$ is the noise contribution to the received pilot sequence of user $k$, and
\begin{equation}
\bar{\bA} =
 \begin{bmatrix}
    \bA_0 & \boldsymbol{\zeta}_0^1 & \hdots &  \boldsymbol{\zeta}_0^{K-1}\\
     \boldsymbol{\zeta}_1^0 & \mathbf{A}_1 & \hdots & \boldsymbol{\zeta}_1^{K-1}\\
     \vdots & \vdots & \ddots & \vdots \\
     \boldsymbol{\zeta}_{K-1}^0 & \boldsymbol{\zeta}_{K-1}^1 & \hdots & \mathbf{A}_{K-1} \\
\end{bmatrix}.
\end{equation}
The matrices $\bA_k$ and $\boldsymbol{\zeta}_{k_1}^{k_2}$ are of size  $N_{\rm p}^k \times L$ and  $N_{\rm p}^{k_1} \times L$, respectively, and have the elements 
 \begin{align} \label{{elementformulaap2}}
    A_{k}(i,l)&= 
    \sum_{q} \sum_{m'} \sum_{n'}  s_{m',n'}^{k}  f[q-l-n'\mbox{$\frac{M}{2}$}]
   f[q-n_i^k\mbox{$\frac{M}{2}$}]  \nonumber\\
  & \times e^{j\frac{2\pi(m'-m_i^k)q}{M}} e^{j\frac{\pi (m'+n'-m_i^k-n_i^k)}{2}} 
   e^{-j\frac{2\pi m' l}{M}},
    \end{align}
    and 
   \begin{align}
    \zeta_{k_1}^{k_2}({i,l})& =  \sum_{q} \sum_{m'} \sum_{n'}  s_{m',n'}^{k_2}  f[q-l-n'\mbox{$\frac{M}{2}$}]
   f[q-n_i^{k_1}\mbox{$\frac{M}{2}$}]  \nonumber\\
  & \times e^{j\frac{2\pi(m'-m_i^{k_1})q}{M}} e^{j\frac{\pi (m'+n'-m_i^{k_1}-n_i^{k_1})}{2}} e^{-j\frac{2\pi m' l}{M}},
    \label{{elementformulazeta}}
    \end{align}
where $(m_i^k,n_i^k)$ is the pair of time-frequency indices that map to the $i$th row of ${\bf A}_k$. We may also note that the columns of ${\bf A}_k$ are aligned with the sample position in the respective channel impulse response ${\bf h}_k$.

The matrix $\bA_k$  is the gain factor indicating the contribution of pilots transmitted by the user $k$ on the respective demodulated signals at the receiver. The matrix $\boldsymbol{\zeta}_{k_1}^{k_2}$, on the other hand, indicates the intrinsic interference produced by the pilots of the user $k_2$ on the demodulated signals of the user $k_1$. Hence, \eqref{eq:linear_MU} decouples the users' channel responses from their training sequences. This enables accurate estimation of all the users' channel impulse responses by taking into account the intrinsic interference between the users' pilot sequences. 

We also note that overlapping of the subcarrier bands in SMT introduces some correlation among the elements of the noise vector $\bar{\boldsymbol{\eta}}$. These  correlations lead to a covariance matrix ${\bf C}_{\bar\eta \bar\eta}$ whose impact on the minimum variance unbiased (MVU) estimate of $\bar{\bf h}$ is reflected in the following equation, \cite{kay1993fundamentals},
\begin{equation}
\hat{\bar{\bh}} = (\bar{\bA}^{\text{H}} \bC_{\bar\eta \bar\eta}^{-1} \bar{\bA})^{-1} \bar{\bA}^{\text{H}} \bC_{\bar\eta \bar\eta}^{-1} \bar{\bz}.
\end{equation}
The $(i,j)$th element of the covariance matrix ${\bf C}_{\bar\eta \bar\eta}$ may be calculated as
\begin{align}
   C_{\bar\eta \bar\eta}(i,j)=&{\rm cov}[\eta_{m_i,n_i},\eta_{m_j,n_j}]\nonumber\\
   =&
     \mathds{E} [\eta_{m_i,n_i}\eta_{m_j,n_j}^*] -  \mathds{E} [\eta_{m_i,n_i}] \mathds{E} [\eta_{m_j,n_j}^*] \nonumber\\
    = & \sigma_{\eta}^2 \sum_{l=-\infty}^{\infty}   f[l] f^*[l]e^{j\frac{2\pi(m_j-m_i)k}{M}}  e^{j\frac{\pi (m_j+n_j-m_i-n_i)}{2}},
\end{align}
where $\sigma_\eta^2$ is the noise variance at the receiver input, and the pairs of $(m_i,n_i)$ and $(m_j,n_j)$ are the pairs of time-frequency indices that map to the $i$th and $j$th elements of $\bar{\boldsymbol{\eta}}$.

It is further noted that the presence of channel noise leads to inaccurate channel estimates  $\hat{\bar{\bh}}$. This adversely affects signal detection. Hence, it is of a great importance to take the noise characteristics into account at the detection stage. To this end, we start from our recent results in \cite{Hoss2006:Spectrally} on channel estimation error where it is shown that the MSE of channel estimation can be found as 
\be
{\rm MSE} = {\rm tr}\{(\bar{\bA}^{\text{H}} \bC_{\bar\eta \bar\eta}^{-1} \bar{\bA})^{-1}\}. 
\ee
This, in part, depends on pilot sets and prototype filter, hence, can be pre-calculated off-line. There is also a proportionality constant equal to the channel noise power that may be added on-line. Accordingly, the estimation error for the channel tap $l$ between BS antenna $i$ and user $k$, $\Delta h_{i,k}[l]$, may be approximated by a complex Gaussian distribution with zero mean and the variance 
\be
\sigma_{\rm et}^2 =  \frac{\rm MSE}{K \times L}. 
\ee
Our proposed equalization techniques in the following section require the statistics of the estimation errors in the frequency domain. Using Parseval's theorem,  one may realize that estimation error at a given subcarrier $m$, also follows complex Gaussian distribution with the variance 
\be
\sigma_{\rm ef}^2=L \sigma_{\rm et}^2. 
\ee
That is, $\Delta H_{m}^{i,k} \sim \mathcal{CN}(0,\sigma^2_{\rm ef})$.

It may be further noted that the limited length of the channel response in the time-domain implies that the estimation errors $\Delta H_{m}^{i,k}$, across different subcarriers, are not independent. Nevertheless, since in this paper, signals from different subcarriers are processed independently, such correlation has no relevant impact on our receiver design and thus is ignored in the rest of our discussions.

\section{Massive MIMO FBMC: Asymptotic Analysis and Equalizer Design}
\label{sec:MAssive_MIMO}
In this section, we develop practical channel equalization techniques for FBMC-based massive MIMO with co-located and distributed antennas. To this end, we start with modifying the existing PDP equalizer in \cite{aminjavaheri2018filter} and propose an equalizer shortening method that leads to a substantially reduced delay. This, in particular, makes our proposal attractive for applications with stringent latency requirements. We also take note that the PDP equalizer of \cite{aminjavaheri2018filter} relies on the assumption that the number of BS antennas is large. We propose a per-subcarrier equivalent channel-based equalization technique that does not rely on this assumption. This new design will be found instrumental in cell-free networks where the number of effective antennas seen by each user terminal remains small and the channel PDPs vary among different receiver antennas. The proposed equivalent channel-based FSE also sets a benchmark for evaluating the performance of the PDP equalizer of \cite{aminjavaheri2018filter} and its modified version here. 

Let us consider a single-cell massive MIMO setup including a BS equipped with $N$ antennas and $K$ single-antenna users. The received signal at a given antenna $i$ can be expressed as 
\be
r_i[l] = \sum_{k=0}^{K-1} x_k[l] \star h_{i,k}[l] + \eta_i[l],
\label{eq:rec_antenna}
\ee
where $x_k[l]$ is user $k$ transmit signal, $\eta_i[l] \sim \mathcal{CN}(0,\sigma^2_{\eta})$ is the additive noise at BS antenna $i$ and $h_{i,k}[l]$ is the channel impulse response between user $k$ and BS antenna $i$. We assume that the BS antenna array is sufficiently compact and model the channels between any given user $k$ and all the BS antennas with the same PDP, i.e., $p_{k}[l]$ for $l=0,\ldots,L-1$. Thus, the channel taps $h_{i,k}[l]$ follow the distribution $\mathcal{CN}(0,p_{k}[l])$ and are independent of one another. {Furthermore, we assume the average transmit power of unity for each user terminal.}

Stacking the demodulated signals corresponding to different BS antennas, after phase adjustment (i.e., removing the phase factor $e^{j\pi(m+n)/2}$) but before taking the real part, into $N \times 1$ vectors $\z_{m,n}$, we have
\be \label{eq:z_mn}
\z_{m,n} = \sum_{n'=-\infty}^{+\infty} \sum_{m'=0}^{M-1} \bH_{mm',nn'} \bs_{m',n'}+\boldsymbol{\eta}_{m,n},
\ee
where the vector $\bs_{m,n}=[s_{m,n}^0,\ldots,s_{m,n}^{K-1}]^{\rm T}$ contains the real-valued data symbols of different users at the time-frequency slot $(m,n)$, $\boldsymbol{\eta}_{m,n}$ is the contribution of noise, and $\bH_{mm',nn'}$ is the $N\times K$ gain matrix among data symbols across both time and frequency with the elements of
 \be \label{eq:H_mmnn}
H_{mm',nn'}^{i,k} = h_{mm'}^{i,k}[n-n'] e^{j(m'+n'-m-n)\frac{\pi}{2}},
\ee
where $h_{mm'}^{i,k}[n] = \big(f_{m'}[l] \star h_{i,k}[l] \star f_m^{*}[l] \big)_{\downarrow \frac{M}{2}}$.

Considering perfect synchronization and knowledge of the channel, and using a per-subcarrier combiner matrix $\bW_m$, the data symbols of different users are estimated as
 \begin{align}\label{eq:estimate_co}
\hat{\bs}_{m,n} = &\Re\big\{\bW_m^{\rm H} \z_{m,n} \big\} \nonumber \\ 
=&\Re\big\{\sum_{n'=-\infty}^{+\infty} \sum_{m'=0}^{M-1} \bG_{mm',nn'} \bs_{m',n'}+\boldsymbol{\eta}'_{m,n}\big\},
\end{align}
where $\bG_{mm',nn'} = \bW_m^{\rm H}\bH_{mm',nn'}$ and $\boldsymbol{\eta}'_{m,n}=\bW_m^{\rm H}\boldsymbol{\eta}_{m,n}$. The common linear combiners, the maximum ratio combining (MRC), the zero forcing (ZF) detector, and the minimum mean square error (MMSE) detector, are respectively defined as 
 \be
 \bW_m = \begin{cases}
\bH_m \bD_m^{-1}, & \text{for MRC,}\\
\bH_m (\bH_m^{\rm H} \bH_m)^{-1}, & \text{for ZF,}\\
\bH_m (\bH_m^{\rm H} \bH_m+\sigma^2_{\eta} \bI_K)^{-1}, & \text{for MMSE},
\end{cases}
\label{eq:combiners}
\ee
where $\bH_m$ is the $N\times K$ channel matrix whose element, $(i,k)$ represents the channel gain at the center of a given subcarrier $m$ between user $k$ and BS antenna $i$, i.e.,  $H_m^{i,k} \triangleq \sum_{l=0}^{L-1} h_{i,k}[l] e^{-j\frac{2 \pi m l}{M}}$. In MRC, the $K \times K $ diagonal matrix $\bD_m$ normalizes the combiner outputs with the coefficients $D_m^{k,k} = \sum _{i=0}^{N-1}|H_m^{i,k}|^2$.

 The authors in \cite{aminjavaheri2018filter} have shown that even with an infinite number of antennas, a residual interference remains at the output of any of the above combiners. 
It has been also noted that for a large number of antennas, all the above combiners converge to $\frac{1}{N} \bH_m$, \cite{ngo2013energy}. 
 Accordingly, the equivalent channel impulse response between the transmitted symbols at subcarrier $m'$ of terminal $k'$ and the received and combined signal at subcarrier $m$ of BS output corresponding to terminal $k$ may be expressed as 
\begin{align}\label{eq:g_{mm'}^{k,k'}[n]}
 g_{mm'}^{k,k'}[n] = \big(f_{m'}[l] \star h_{k,k',m}^{\rm (eqvlt)}[l] \star f_m^*[l]  \big)_{\downarrow \frac{M}{2}}.
\end{align}
where
\be\label{eq:equivalent-channel}
h_{k,k',m}^{\rm (eqvlt)}[l]=\frac{1}{N}\sum_{i=0}^{N-1}(H_m^{i,k})^*h_{i,k'}[l],
\ee
is the combined/equivalent channel between the user terminal $k'$ and the combiner output of the $k$th user over the subcarrier band $m$.

In \cite{aminjavaheri2018filter}, it is shown that for a large number of antennas $N$, at the BS,  $h_{k,k',m}^{\rm (eqvlt)}[l]$ vanishes to zero, when $k\ne k'$, and when $k=k'$, $h_{k,k,m}^{\rm (eqvlt)}[l]$ in \eqref{eq:g_{mm'}^{k,k'}[n]} may be replaced by
\be\label{eq:pmk[l]}
\bar{p}_{m,k}[l]= p_{k}[l]   e^{j2\pi lm/M},
\ee
where $p_{k}[l]$ is the channel PDP between the user terminal $k$ and the BS antennas.

It has been further argued in \cite{aminjavaheri2018filter} that the presence of $h_{k,k',m}^{\rm (eqvlt)}[l]$ (equivalently, $\bar{p}_{m,k}[l]\delta_{kk'}$) in \eqref{eq:g_{mm'}^{k,k'}[n]} breaks the Nyquist property between the transmit and receive prototype filters. Hence, a per subcarrier  equalizer should be adopted to undo the effect of $h_{k,k',m}^{\rm (eqvlt)}[l]$. Accordingly, \cite{aminjavaheri2018filter} has proposed the zero forcing equalizer 
\be\label{eq:Phimk}
 \Phi_{m,k}(\omega)=1/\bar{P}_{m,k}(\omega),
 \ee 
where $\bar{P}_{m,k}(\omega)$ is the discrete-time Fourier transform (DTFT) of $\bar{p}_{m,k}[l]$. Furthermore, it has been noted in \cite{aminjavaheri2018filter} that the equalizer introduced in \eqref{eq:Phimk} is a modulated version of a baseband equalizer common to all the subcarriers. Hence, a single equalizer design will suffice. 

The rest of this section is organized as follows. We first revisit the PDP-based equalizer design of \cite{aminjavaheri2018filter}, take note that it is only applicable to large and co-located antenna systems, mention some of the inherent problems of this design, and propose a new design technique for resolving these problems. Next, we note that in distributed antenna systems, where the antennas that are attached to a BS are distributed over a wide area for a higher diversity gain, \cite{ngo2017cell}, the underlying channels have different PDPs and thus the PDP equalizer design is not applicable. The same is true for cell-free MIMO systems \cite{bjornson2019making}. Moreover, we note that when the number of BS antennas is small, the PDP equalizer may not be effective. We thus propose a new equalizer design that can serve these scenarios.  Lastly, we take note that the distributed antenna/cell-free systems need a special treatment to keep the same fairness for all users in the network. 

\subsection{Large  and co-located antenna systems}\label{subsec:PDP_colocated}
Further study of the above equalizer design reveals that, \cite{aminjavaheri2018filter}, the equalizer \eqref{eq:Phimk} may be implemented at baseband, i.e., after the analysis filter bank, the $M/2$-fold decimation, and the combining \eqref{eq:estimate_co}, but before taking the real-part. This, obviously, reduces the complexity of the receiver significantly as the equalizer is implemented at a reduced sampling rate. Moreover, \cite{aminjavaheri2018filter} has noted that only a common equalizer may be used for each subcarrier, as against $N$ separate equalizers (one for each antenna). The implemented equalizer in the baseband is the demodulated and $M/2$-fold decimated version of $\Phi_{m,k}(\omega)$, in the time domain. Furthermore, the demodulated version of $\Phi_{m,k}(\omega)$, i.e., $\Phi_{0,k}(\omega)$, before decimation has to be band limited through an antialiasing filter. Another point that needs to be noted is that the baseband equalizer designed here is an FSE \cite{1457566} with tap-spacing of half symbol spacing.

In typical FBMC systems, the prototype filter is often chosen to be a square-root Nyquist filter with an excess bandwidth of 100\%, equivalent to a roll-off factor $\alpha=1$. This leaves no room for the transition band of the antialiasing filter that was mentioned above. Hence, \cite{aminjavaheri2018filter} has proposed the use of a brick-wall antialiasing filter. Such brick-wall filter, unfortunately, increases the length of the equalizer significantly. This leads to both complexity and latency issues. Here, we propose a modified design that solves these shortcomings of the PDP equalizer design of \cite{aminjavaheri2018filter}. 

We first note that for $k=k'$ and $m=m'$, replacement of \eqref{eq:pmk[l]} in \eqref{eq:g_{mm'}^{k,k'}[n]} leads to
\be\label{eq:g_{mm}^{k,k}[n]}
 g_{mm}^{k,k}[n] = \big(f_{m}[l] \star  p_{k}[l]   e^{j2\pi lm/M}\star f_m^*[l]  \big)_{\downarrow \frac{M}{2}},
\ee
where $f_m[l] \triangleq f[l] e^{j\frac{2 \pi m l}{M}}$. Noting that $f_m[l]$ is a bandpass filter with a bandwidth of $2/M$, centered at the subcarrier frequency $f_c=1/M$, and following the basic theory of multirate signal processing \cite{farhang2008signal,vaidyanathan2006multirate}, it is straightforward to show that $g_{mm}^{k,k}[n]$ is a baseband signal that spans over the frequency band $-\frac 1M<f<\frac 1M$. It is also known that the decimation by $M/2$ may be viewed as a demodulation process that removes the modulation factor $e^{j2\pi lm/M}$ from all the terms on the right-hand side of \eqref{eq:g_{mm}^{k,k}[n]}, leading to 
\be\label{eq:g_k[n]}
  g_k[n]= \big(f[l] \star  p_{k}[l]   \star f[l]  \big)_{\downarrow \frac{M}{2}},
\ee
where we have defined the pulse-shape $g_k[n]=g_{mm}^{k,k}[n]$, noting that the right-hand side of \eqref{eq:g_k[n]} is independent of $m$. 

The result in \eqref{eq:g_k[n]} shows that a common equalizer that removes any inter-symbol interference (ISI) generated by the pulse-shape $g_k[n]$ may be designed and applied to all subcarrier signals at the FBMC receiver output, i.e., after analysis filter bank, decimation, and combining. This equalizer is a fractionally spaced one, \cite{farhang2008signal,farhang2013adaptive}, that, for any $m$, covers  the $m$-th band of the filter bank, including the portions of the band that overlap with the adjacent bands. It thus also removes the intrinsic interference from the adjacent bands. The equalizer design, here, may be a ZF or an MMSE one that can provide a satisfactory performance with a very small number of taps. As a result, our proposal here addresses the aforementioned complexity and latency issues of the equalizer in \cite{aminjavaheri2018filter}. Numerical examples that show the impact of these modifications on the design of the equalizer, when compared to equalizer design of \cite{aminjavaheri2018filter}, are discussed in Section~\ref{sec:simulation}.

\subsection{Small antenna systems}

We first note that when the number of BS antennas is small, the ZF or MMSE combiners significantly outperform MRC, hence, should be adopted. When any of these combiners is adopted, the equivalent channel response \eqref{eq:equivalent-channel}, for $k=k'$, should be replaced by
\be\label{eq:equivalent-channel-general}
h_{k,k,m}^{\rm (eqvlt)}[l]=\sum_{i=0}^{N-1}(W_m^{i,k})^*h_{i,k}[l],
\ee
where the coefficients $W_m^{i,k}$ are the combiner coefficients given by the ZF or MMSE combiner in \eqref{eq:combiners}. Substituting \eqref{eq:equivalent-channel-general} in \eqref{eq:g_{mm'}^{k,k'}[n]} leads to
\be\label{eq:g_{mm}^{k,k}[n]2}
 g_{mm}^{k,k}[n] = \left[f_{m}[l] \star \left(\sum_{i=0}^{N-1}(W_m^{i,k})^*h_{i,k}[l]\right) \star f_m^*[l]  \right]_{\downarrow \frac{M}{2}}.
\ee
Following the discussions surrounding equations \eqref{eq:g_{mm}^{k,k}[n]} and \eqref{eq:g_k[n]}, above, one will find that, here, $g_{mm}^{k,k}[n]$ is a baseband pulse-shape that, unlike $g_k[n]$ in \eqref{eq:g_k[n]}, varies with the subcarrier index $m$. 
Hence, here, for each subcarrier band, we propose designing a ZF or MMSE equalizer that removes ISI in the pulse-shape  $g_{mm}^{k,k}[n]$ of \eqref{eq:g_{mm}^{k,k}[n]2}.  The numerical results that are presented in Section~\ref{sec:simulation} show an excellent performance of this design. In particular, we find that even for a small number of BS antennas, when this equalizer design is applied to the output of a ZF or MMSE combiner, the receiver performance remains very similar to that of the single user performance. This observation may be explained as follows. The ZF or MMSE combiner removes most of the multi-user interference. The remaining distortion, which is limited to the user of interest, is then removed by the equalizer.

\subsection{Cell-free/distributed antenna systems}
\label{sec:system_model}
\begin{figure}
		\centering
		\includegraphics[scale=3.2,trim={0 0 0 0},clip]{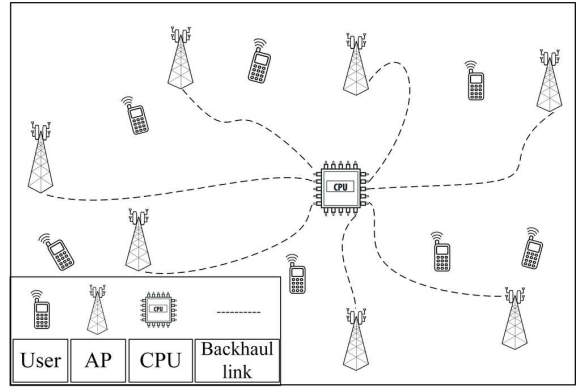}
	\caption{Cell-free massive MIMO network architecture.}
		\label{fig:cell_free_diagram}
		\vspace{-0.4 cm}
	\end{figure}

In this scenario, $N_{\rm AP}$  distributed access points (APs) are connected to a central processing unit (CPU) through a backhaul network as in Fig \ref{fig:cell_free_diagram}. We consider a centralized processing scenario in the uplink, similar to \cite{bjornson2019making}, where the CPU performs detection by processing all the received signals from $N$ total antennas. It is worth mentioning that distinct APs are assigned antenna indices that belong to mutually disjoint subsets of the available antennas, i.e., a given antenna is only assigned to one AP. In this work, without loss of generality, we consider APs that are equipped with an equal number of antennas, i.e., $\frac{N}{N_{\rm AP}}$ antennas each. 

In the previous cases, we assumed perfect power control. Hence, we assumed the normalized PDP which would result in $\sum_l p_{i,k}[l]=1$, for all pairs of $i$ and $k$. In a cell-free/distributed antenna scenario, $\sum_l p_{i,k}[l]=\beta_{i,k}$, where $\beta_{i,k}$ is a large-scale fading coefficient that depends on the distance between a given user $k$ and an antenna $i$ and any shadowing effects,\cite{ngo2017cell,nayebi2015cell}.  We assume the same normalized PDP and the large-scale fading coefficient for a given user and all the antennas of a given AP. 

    \begin{figure*}[t]
		\centering
		\includegraphics[scale=0.4,trim={0 0 0 0},clip]{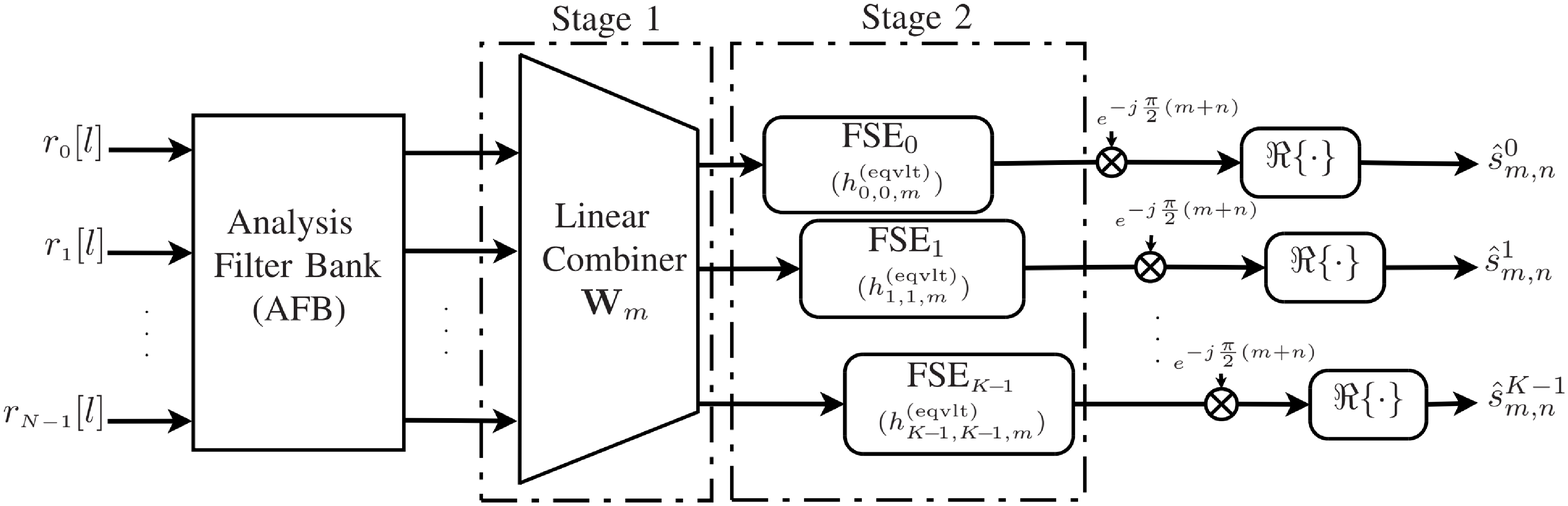}
		\caption{The proposed two-stage equalization receiver structure. The first stage is a conventional linear combining that equalizes the channels at the center of the subcarrier bands and separates different users' signals. At the second stage, the output of the first stage is passed through a set of FSEs that repeat for a given subcarrier $m$ and a given user $k$ for removal of any residual ISI and ICI. 
		}
		\label{blockdiagram}
	\end{figure*}

The presence of different large-scale fading coefficients with large variations between a given user and APs in a cell-free massive MIMO architecture results in fairness issues \cite{ngo2017cell}. This necessitates the application of effective power control methods that can strike a balance between fairness and average SINR performance \cite{simonsson2008uplink}. Hence, here, we deploy the fractional power control that is proposed for OFDM-based cell-free massive MIMO in \cite{nikbakht2020uplink}. This is an extended form of the power control method used in the long term evolution (LTE) standard. Based on the results of \cite{nikbakht2020uplink}, the power control coefficient for a given user $k$ in the uplink of a cell-free massive MIMO can be obtained as
   \be
\mu_k \propto \frac{1}{(\sum_{i=0}^{N-1}\beta_{i,k})^{\nu}}.
\ee
where $\nu$ is a design parameter to be set between $0.5$ and $0.7$,  \cite{simonsson2008uplink,yates1995framework}. By changing $\nu$, we are able to adjust the trade-off between fairness and average SINR. Thus, the transmit signal of user $k$ may be expressed as
\be
x_k[l] = \sum_{m=0}^{M-1} \sum_{n=-\infty}^{ \infty} \sqrt{\mu_k} s_{m,n}^k f_{m,n}[l].
\label{bbeq_cell_free}
\ee
This is a simple modification to \eqref{bbeq} which leads to the following modified form of the equivalent channel \eqref{eq:equivalent-channel-general}.
\be\label{eq:equivalent-cell-free-channel}
h_{k,k,m}^{\rm (eqvlt)}[l]=\sum_{i=0}^{N-1}(W_m^{i,k})^*\sqrt{\mu_k}h_{i,k}[l].
\ee
The FSE design thus follows accordingly.

\subsection{Summary}
Fig.~\ref{blockdiagram} summarizes our proposal in this section as a two-stage equalization process following the analysis filter bank (AFB) steps expressed by \eqref{eq:z_mn} and \eqref{eq:H_mmnn}. The first stage is a conventional linear combining that equalizes the channels at the center of the subcarrier bands and separates different users' signals. This stage can be thought of as a coarse equalization and multiuser detection stage. In the second stage, the output of the first stage is passed through a set of FSEs that repeat for each subcarrier $m$ and every user $k$ for removal of any residual ISI as well as any inter-carrier interference (ICI).

\section{Channel Equalization with Imperfect CSI }
\label{sec:imperfect}
The equalizer designs that have been proposed so far are based on  the assumption of having perfect CSI at all the BS antennas. Obviously, in practical systems, the presence of channel estimation errors  can adversely affect the performance of these equalizers. Here, we use the results of Section~\ref{sec:estimation} and modify our proposed designs in Section~\ref{sec:MAssive_MIMO} to take the statistics of the channel estimation errors into account in the detection stage.


\subsection{Large  and co-located antenna systems}
The combiners with imperfect CSI are derived by substituting $H^{i,k}_{m}$ with $\hat{H}^{i,k}_{m} = H^{i,k}_{m} + \Delta H^{i,k}_{m}$ in \eqref{eq:combiners}. Here, to get some insight, we look at these combiners by exploring their performance in the asymptotic regime where $N$ grows to a large value. 

We recall that, in the case of MRC, $\bD_m$  is a diagonal matrix with the diagonal elements $D_m^{k,k} = \sum_{i=0}^{N-1}|H_m^{i,k}|^2$. With imperfect CSI, this becomes
\be
\hat{D}_m^{k,k} = \sum_{i=0}^{N-1}|H_m^{i,k}+\Delta H_m^{i,k}|^2.
\ee
Assuming  uncorrelated estimation errors and channel gains, by the law of large numbers, in the asymptotic regime, $\hat{D}_m^{k,k}$ converges to
\be
 N\mathds{E}\{|H_m^{i,k}|^2\}+N\mathds{E}\{|\Delta H_m^{i,k}|^2\}=N+N\sigma_{\rm ef}^2.
\ee

Using a similar approach for the  ZF combiner, it is not hard to show that in the asymptotic regime $\hat{\bH}_m^{\rm H} \hat{\bH}_m$ converges to $\hat\bD_m$. Hence, the ZF combiner performance loss due to channel estimation error follows that of the MRC.  

In the MMSE combiner case, we argue that as $N$ grows large, $\sigma_{\eta}^2$ will become  negligible when compared to $N+N \sigma_{\rm ef}^2$ and, hence, MMSE combiner will converge to the ZF combiner which in the asymptotic regime is similar to the MRC. These show that in the asymptotic regime, all three combiners converge to $\hat{\bW}_m = \frac{1}{N(1+\sigma_{\rm ef}^2)} \hat{\bH}_m$. Consequently, from \eqref{eq:estimate_co}, the combined/equivalent channel between the transmit symbol at subcarrier $m'$ of terminal $k'$ and the received one at subcarrier $m$, at the $k$th combiner output converges to
\begin{align}
        h_{k,k',m}^{\rm (eqvlt)}[l] =&\frac{1}{N(1+\sigma_{\rm ef}^2)}\sum_{i=0}^{N-1}  (\hat{H}_m^{i,k})^*  h_{i,k'}[l] . \label{eq:equivalent_channel_IP}
\end{align}
Moreover, for large values of $N$, \eqref{eq:equivalent_channel_IP}, reduces to
\begin{equation}
h_{k,k',m}^{\rm (eqvlt)}[l] = \frac{1}{1+\sigma_{\rm ef}^2}    \mathds{E} \big\{ (\hat{H}_m^{i,k})^*  h_{i,k'}[l] \big\}.
\end{equation}
Assuming independent channel responses for different users, independent channel taps, and uncorrelated channel estimation errors, one will find that
\be \label{eq:expectationpdp}
\mathds{E}\big[ (\hat{H}_m^{i,k})^* h_{i,k'} \big] = p_{k}[l] e^{j2\pi lm/M} \delta_{kk'}.
\ee
Thus, the equivalent channel converges to
\be
h_{k,k',m}^{\rm (eqvlt)}[l]  =  \tilde{p}_{m,k}[l]\delta_{kk'},
\label{eq:eq_channel_imperfect_co}
\ee
where
\be 
\begin{aligned}\label{eq:ColocatedPDP_imperfect}
&\tilde{p}_{m,k}[l]= \frac{\bar{p}_{m,k}[l]}{1+\sigma_{\rm ef}^2}.
\end{aligned}
\ee

The above results lead to the following conclusion. To compensate for the imperfect CSI, the scaling factor $1/({1+\sigma_{\rm ef}^2})$ should be added to the PDP $\bar{p}_{m,k}[l]$. This is equivalent to adding the {\em correction factor} $1+\sigma_{\rm ef}^2$ to the designed equalizer.   The accuracy of this modified design is corroborated through simulations in Section~\ref{sec:simulation}.

\vspace{3mm}

\noindent
{\em PDP approximation:}
In the imperfect CSI scenario, the knowledge of channel statistics is not always available. Thus, we propose to approximate the PDP by taking the average 
\be \label{eq:pdp_approx}
\hat{p}_k[l]=\frac{1}{N} \sum_{i=0}^{N-1}|\hat{h}_{i,k}[l]|^2.
\ee
This is then replaced for $p_k[l]$ in \eqref{eq:expectationpdp}.


\subsection{Small antenna systems}
Substituting the channel estimates in \eqref{eq:equivalent-channel-general}, we  obtain  the equivalent channel estimate
 \be\label{eq:ResCh_MMSE}
 \hat{h}_{k,k',m}^{\rm (eqvlt)}[l]  = \sum_{i=0}^{N-1} (\hat{W}_m^{i,k})^* \hat{h}_{i,k'}[l],
\ee
where $ \hat{h}_{i,k'}[l]= h_{i,k'}[l]+ \Delta h_{i,k'}[l]$.

Here, for moderate and large values of $N$,
 \be
\begin{aligned}\label{eq:MMSE_EqCh}
 &\hat{h}_{k,k',m}^{\rm (eqvlt)}[l]  \! \rightarrow \! \frac{1}{1+\sigma_{\rm ef}^2}  \mathds{E} \big\{ (\hat{H}_m^{i,k})^*  \hat{h}_{i,k'}[l] \big\}.
 \end{aligned}
\ee
Assuming independent channel responses and uncorrelated estimation errors,  
\be
\begin{aligned}
\mathds{E} \big\{ (\hat{H}_m^{i,k})^*  \hat{h}_{i,k'}[l] \big\}=& \sum_{l'=0}^{L-1} \mathds{E}\big\{{h}^*_{i,k'}[l'] {h}_{i,k}[l] \big\} e^{j2\pi l'm/M} \\ & \!+\!\sum_{l'=0}^{L-1} \mathds{E}\big\{\Delta h_{i,k}^*[l'] \Delta h_{i,k'}[l] \big\} e^{j2\pi l'm/M} \\
= & ~\bar{p}_{m,k}[l]\delta_{kk'}+\sigma_{\rm et}^2 e^{j2\pi lm/M} \delta_{kk'}.
\end{aligned}
\ee
Hence, \eqref{eq:MMSE_EqCh} can be written as
\be\label{eq:MMSEinput_est}
\hat{h}_{k,k',m}^{\rm (eqvlt)}[l] \rightarrow h_{k,k',m}^{\rm (eqvlt)}[l] + \frac{\sigma_{\rm et}^2 \delta_{kk'}}{1+\sigma_{\rm ef}^2}e^{j2\pi lm/M}.
\ee
where the second term on the right-hand side of \eqref{eq:MMSEinput_est} is due to the effect of channel estimation errors. This can be simply mitigated by subtracting the {\em correction term} $\frac{\sigma_{\rm et}^2 \delta_{kk'}}{1+\sigma_{\rm ef}^2}e^{j2\pi lm/M}$ from the equivalent channel estimate \eqref{eq:ResCh_MMSE} and designing the equalizer based on the modified channel estimate. The efficacy of this solution is corroborated by simulations in Section~\ref{sec:simulation}.

\subsection{Cell-free/distributed antenna systems}
Here, \eqref{eq:ResCh_MMSE} is replaced by
 \be\label{eq:cell_ResCh_MMSE}
 \hat{h}_{k,k',m}^{\rm (eqvlt)}[l]  = \sum_{i=0}^{N-1} (\hat{W}_m^{i,k})^* \sqrt{\mu_k}\hat{h}_{i,k'}[l],
\ee

Following the same line of derivations as in \cite{ngo2013energy}, as the number of antennas, $N$, grows large, here, the combiners converge to $\bW_m={\rm diag}([\sum_{i=0}^{N-1}\beta_{i,0}, ..., \sum_{i=0}^{N-1}\beta_{i,K-1}])^{-1} \bH_m$. Hence, 
\begin{align}\label{eq:cell_h_eqvlt}
 &\hat{h}_{k,k',m}^{\rm (eqvlt)}[l] \! \rightarrow \! \frac{\sqrt{\mu_k}}{\sum_{i=0}^{N-1}\beta_{i,k}+N\sigma_{\rm ef}^2} \sum_{i=0}^{N-1}  \mathds{E} \big\{ (\hat{H}_m^{i,k})^*  \hat{h}_{i,k'}[l] \big\}  .
 \end{align}
Also, assuming independent channel responses and uncorrelated estimation errors, one will find that 
\begin{align} \label{eq:MMSE_EqCh_cell}
&\mathds{E} \big\{ (\hat{H}_m^{i,k})^*  \hat{h}_{i,k'}[l] \big\}= \sum_{l'=0}^{L-1} \mathds{E}\big\{{h}^*_{i,k'}[l'] {h}_{i,k}[l] \big\} e^{j2\pi l'm/M} \nonumber\\ &~~~~~~~+\sum_{l'=0}^{L-1} \mathds{E}\big\{\Delta h_{i,k}^*[l'] \Delta h_{i,k'}[l] \big\} e^{j2\pi l'm/M} \nonumber\\
&~~~~~~~= p_{i,k}[l]e^{j2\pi lm/M} \delta_{kk'}+\sigma_{\rm et}^2 e^{j2\pi lm/M} \delta_{kk'}.
\end{align}
Substituting \eqref{eq:MMSE_EqCh_cell} in \eqref{eq:cell_h_eqvlt}, leads to 
\begin{align}\label{eq:MMSEinput_est_cell}
&\hat{h}_{k,k',m}^{\rm (eqvlt)}[l]\rightarrow h_{k,k',m}^{\rm (eqvlt)}[l] \! + \frac{N\sigma_{\rm et}^2 \sqrt{\mu_k} \delta_{kk'}}{\sum_{i=0}^{N-1} \beta_{i,k}+N\sigma_{\rm ef}^2}e^{j2\pi lm/M} .
 \end{align}

Similar to the co-located setup in the previous subsection, here, subtracting the {\em correction factor} $ \frac{N\sigma_{\rm et}^2 \sqrt{\mu_k} \delta_{kk'}}{\sum_{i=0}^{N-1} \beta_{i,k}+N\sigma_{\rm ef}^2}e^{j2\pi lm/M} $ from the equivalent channel estimate \eqref{eq:cell_ResCh_MMSE} and designing the equalizer based on the modified channel estimate, mitigates the channel estimation error effects. This statement is confirmed through computer simulations in the following section.

 \section{Simulation Results}
 \label{sec:simulation}
 In this section, we evaluate our mathematical developments throughout the paper by computer simulations. We first present a set of results for a single cell scenario with co-located antennas at the BS. Then, simulation results that evaluate the performance of the proposed methods in a cell-free scenario are presented.
  
\subsection{Single cell scenario}
 We consider QAM (quadrature amplitude modulation) symbols to be transmitted over $M=64$ subcarriers of the SMT system with PHYDYAS prototype filter, \cite{bellanger2010fbmc}, and overlapping factor $\kappa=4$. In our simulations, we use tap delay line-C (TDL-C) 5G channel model, \cite{etsi2017138}. This model provides a PDP based on a normalized root mean square (RMS) delay spread. Following the instructions in \cite{etsi2017138}, we randomly scale the normalized RMS delay spreads for different users in each simulation instance using a uniform distribution to achieve the RMS delay spreads within the range  $[90~{\rm ns},110~{\rm ns}]$, i.e., for channels with moderate lengths. The reason for this is that in realistic scenarios, PDPs between the users and the BS antennas are different. Perfect power control is assumed and, thus, the PDPs are normalized, i.e., we let $\sum_{l=0}^{L-1}p_k[l]=1$ for $k=0,\ldots, K-1$. We consider the input signal-to-noise ratio (SNR) of $10$~dB at the BS antennas unless otherwise is stated. We set the sampling frequency to $15.36$~MHz. This leads to the subcarrier spacing of $240$~kHz, which is inline with 5G NR specifications, \cite{etsi2017138}. We have obtained our results for 1000 independent realizations of the channel with $K=4$ users. 

\begin{figure}[t]
		\centering
		\includegraphics[scale=0.65,trim={0 0 0 0},clip]{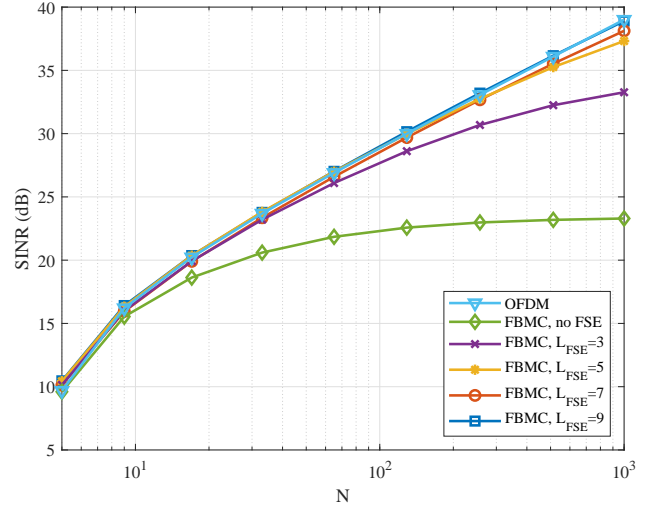}
		\caption{SINR vs. the number of BS antennas, $N$. The BS antennas are co-located and FSE design is based on the PDP of the underlying channels. Different choices of $L_{\rm FSE}$ are examined. OFDM results are presented as a benchmark. 
		}
		\label{fig:SINRLength}
	\end{figure}

The URLLC applications require the minimum possible delay. Thus, finding a minimum acceptable equalizer length is of a great importance. Fig.~\ref{fig:SINRLength} shows the SINR performance of the PDP-based FSE design as $L_{\rm FSE}$ and the number of base station antennas, $N$, vary. The result when the FSE is absent is also presented. Our results show the efficacy of the proposed FSE technique in removing the SINR saturation problem of the single stage equalization, i.e., the one with linear combining only.  We also note that since in OFDM the channel is frequency flat over each subcarrier band, OFDM may be used as a benchmark for evaluating the efficacy of the proposed two-stage equalization. For the first stage, we consider ZF. For the channel scenario that is studied here, we find that an FSE length of $L_{\rm FSE}=3$ leads to a significant improvement over the case where there is no FSE. The improvement approaches the performance of OFDM for $L_{\rm FSE}=9$ and a minor deviation from this optimal performance is observed as $L_{\rm FSE}$ decreases to the values of $7$ and $5$. It appears that in the present channel scenario $L_{\rm FSE}=5$ strikes a good balance between the receiver complexity and latency, as well as its performance. Examining other channel models, we have found that this compromise choice remains the same over a wide range of channel conditions. Compared to the FSE design presented in \cite{aminjavaheri2018filter}, this FSE length is an order of magnitude smaller. The need for a very long FSE in \cite{aminjavaheri2018filter} was an outcome of using the brick-wall antialiasing filter that was mentioned in Section~\ref{sec:MAssive_MIMO}.

\begin{figure}[t]
		\centering
		\includegraphics[scale=0.65,trim={0 0 0 0},clip]{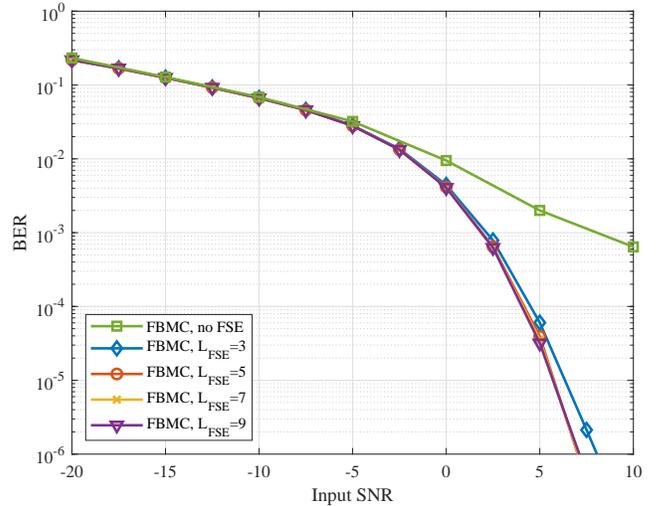}
		\caption{BER vs. input SNR, for $N=100$. The BS antennas are co-located and FSE design is based on the PDP of the underlying channels. Different choices of $L_{\rm FSE}$ are examined. 
		}
		\label{fig:BERLength}
	\end{figure}

In Fig.~\ref{fig:BERLength}, we compare the uncoded bit error rate (BER) performance of the FSE with different lengths in a scenario with $N=100$ BS antennas and the constellation size of $64$-QAM. From Fig.~\ref{fig:BERLength}, one may realize that using the proposed FSE brings a significant BER performance improvement even when $L_{\rm FSE}=3$. It is worth noting that for SNRs up to $5$~dB, the BER performance when $L_{\rm FSE}=3$ is very close to those with a longer $L_{\rm FSE}$. At higher SNRs, this short equalizer leads to a loss of less than  $1$~dB at the BER of $10^{-6}$. Based on these results, in the following experiments, we set $L_{\rm FSE}=5$.

\begin{figure}[t]
		\centering
		\includegraphics[scale=0.65,trim={0 0 0 0},clip]{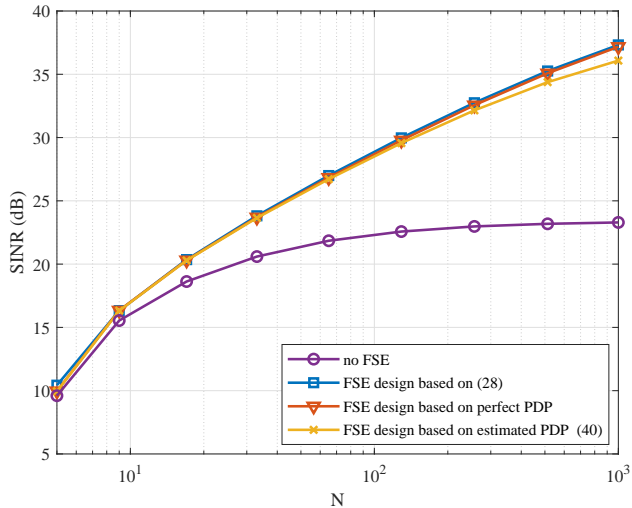}
		\caption{Output SINR vs. the number of BS antennas, $N$, for different FSE designs.  Perfect CSI is assumed and $L_{\rm FSE}=5$. }
		\label{fig:SINRperfect}
	\end{figure}

In Fig.~\ref{fig:SINRperfect}, we evaluate the performance of our proposed FSE with the length $L_{\rm FSE}=5$ for different designs. The designs that are presented are: (i) based on the equivalent channel (\ref{eq:equivalent-channel-general}); (ii) based on the exact PDP that has been used to generate the random channels; and (iii) based on the approximate/estimated PDP (\ref{eq:pdp_approx}). It is also assumed that the channel estimates are perfect.  These results show that the proposed FSE using the equivalent channel and the exact PDP  lead to about the same performance. This is while the proposed FSE with the approximate/estimated PDP leads to a negligible performance loss. This loss is only observable for a large number of BS antennas where the output SINR approaches a large value and noise effects are no longer dominant. These results show that when the second order statistics of the channel is available at the BS, the computational burden for calculation of the exact equivalent channel that needs to be treated separately at different subcarriers can be avoided.

\begin{figure}[t]
		\centering
		\includegraphics[scale=0.65,trim={0 0 0 0},clip]{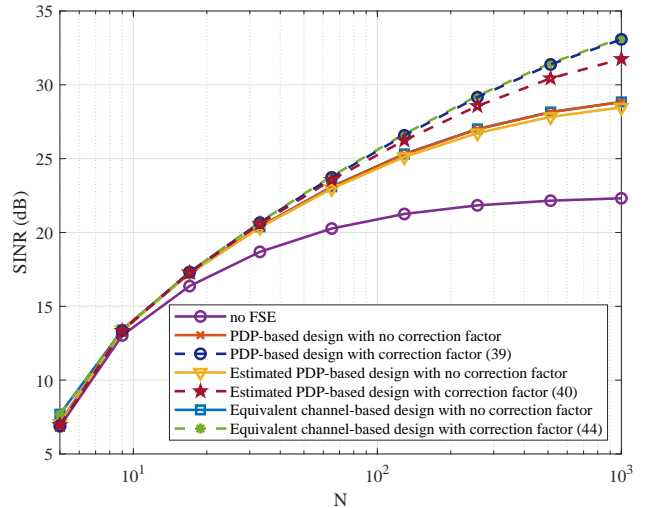}
		\caption{Output SINR vs. the number of BS antennas, $N$, for different FSE designs.  $L_{\rm FSE}=5$. }
		\label{fig:SINRimperfecet}
	\end{figure}

To study the imperfect CSI scenario, our proposed channel estimation method in Section~\ref{sec:estimation} is deployed. We insert $\kappa-1 = 3$ guard symbols
in time to isolate the preamble from the data symbols. Considering the imperfect CSI statistics, in Fig.~\ref{fig:SINRimperfecet}, we study the efficacy of the proposed modified equalizers in Section~\ref{sec:imperfect}. These results show that the modified equalizers lead to an improved performance compared with the ones in Section~\ref{sec:MAssive_MIMO} that do not take into account the imperfect CSI effects. According to the results of Fig.~\ref{fig:SINRimperfecet}, as the number of BS antennas increases, imperfect CSI effects become more problematic if not compensated. 

\begin{figure}[t]
		\centering
		\includegraphics[scale=0.65,trim={0 0 0 0},clip]{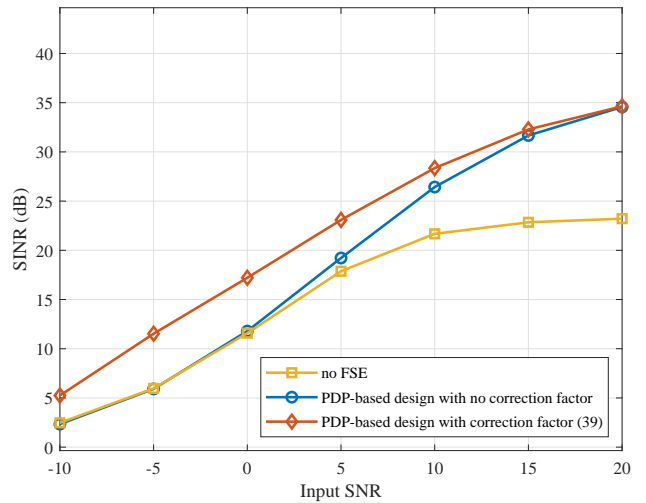}
		\caption{Output SINR vs. input SNR.
		}
		\label{fig:SINRsweep}
	\end{figure}

Noting that channel estimation errors leading to imperfect CSI depend on the noise level at the input of BS antennas, in Fig.~\ref{fig:SINRsweep}, we use estimated channels and study the output SINR as a function of the input SNR for $N=200$. Our results in this figure show that as the input SNR increases, hence, noise level decreases, channel estimates become more accurate, and thus, the equalizers of Section~\ref{sec:MAssive_MIMO} achieve a similar performance to the modified ones in Section~\ref{sec:imperfect}. On the other hand, at lower values of SNR, the proposed modifications in Section~\ref{sec:imperfect} can lead up to $6$~dB SINR improvement. Last but not least, while the single-stage equalization (i.e., linear combining only) leads to about the same performance as the two-stage equalization in the low SNR regime, the addition of the FSE (the second stage) can lead to a gain of $10$~dB or more at higher values of SNR.

 \subsection{Cell-free scenario}
 \label{sec:cellfresimulation}
We perform  simulations for a cell-free massive MIMO setup with APs that are located on a regular grid in a $2\times2$ square kilometers area. Each AP has $4$ antennas. We deploy the wrap-around technique of \cite{bjornson2019making} to imitate an infinite area and thus, avoid boundary effects. We consider $K=4$ users that take random locations in each realization. We consider the same PDPs as the co-located setup. However, to take into account the distribution of antennas/APs,  large-scale fading coefficients are added to different channels. The large-scale fading coefficients are modeled according to the COST Hata model as \cite{damosso1999cost}
\be
10\text{log}_{10}(\beta_{i,k}) = -135-35\text{log}_{10}(d_{i,k})-\mathcal{X}_{i,k},
\ee 
where $d_{i,k}>10~{\rm m}$ is the distance between a given user $k$ and antenna $i$ in kilometers and $\mathcal{X}_{i,k} \sim \mathcal{CN}(0, \sigma^2_{\mathcal{X}})$  represents shadowing effect with $\sigma^2_{\mathcal{X}}=8$ dB. Variance of noise is calculated using the noise figure as $\sigma^2_{\eta} = \mathcal{K} \times \kappa_{\rm B} \times  B \times {\rm NF}$,
where $\mathcal{K}$, $\kappa_{\rm B}$, $B$, and NF are temperature in kelvin, Boltzmann constant, bandwidth, and noise figure, respectively. Here, we let $\mathcal{K}=290$~K, $\kappa_{\rm B} = 1.3 \times 10^{-23}$~J/K, $B=20$~MHz and NF~$=9$ dB. The maximum transmit power of each user is assumed to be $200$~mW.

\begin{figure}
		\centering
		\includegraphics[scale=0.625,trim={0 0 0 0},clip]{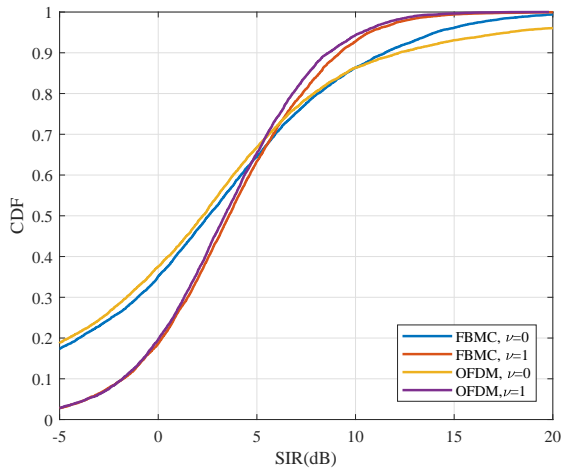}
		\caption{Comparison of the empirical CDF of OFDM and FBMC while using maximum power and power control.}
		\label{fig:cdf}
	\end{figure}

Fig. \ref{fig:cdf} illustrates the cumulative distribution function (CDF) of the signal-to-interference ratio (SIR) performance for FBMC- and OFDM-based cell-free MIMO setup having $9$ APs in an area of $2\times 2$ km and $4$ users with and without power control, i.e., $\nu=0$ and $\nu=1$, respectively. The results show that power control leads to more stable values (i.e., less variation) in SIRs. This is inline with the previous results on OFDM in the literature \cite{nikbakht2020uplink}. Our results, here, confirm that the same is true for FBMC, and power control has almost the same impact on both OFDM and  FBMC. Following the recommendations made in the literature,  \cite{simonsson2008uplink,yates1995framework} and \cite{nikbakht2020uplink}, in the rest of this section, we consider the fractional power control with $\nu=0.5$. 

	\begin{figure}
		\centering
		\includegraphics[scale=0.61,trim={0 0 0 0},clip]{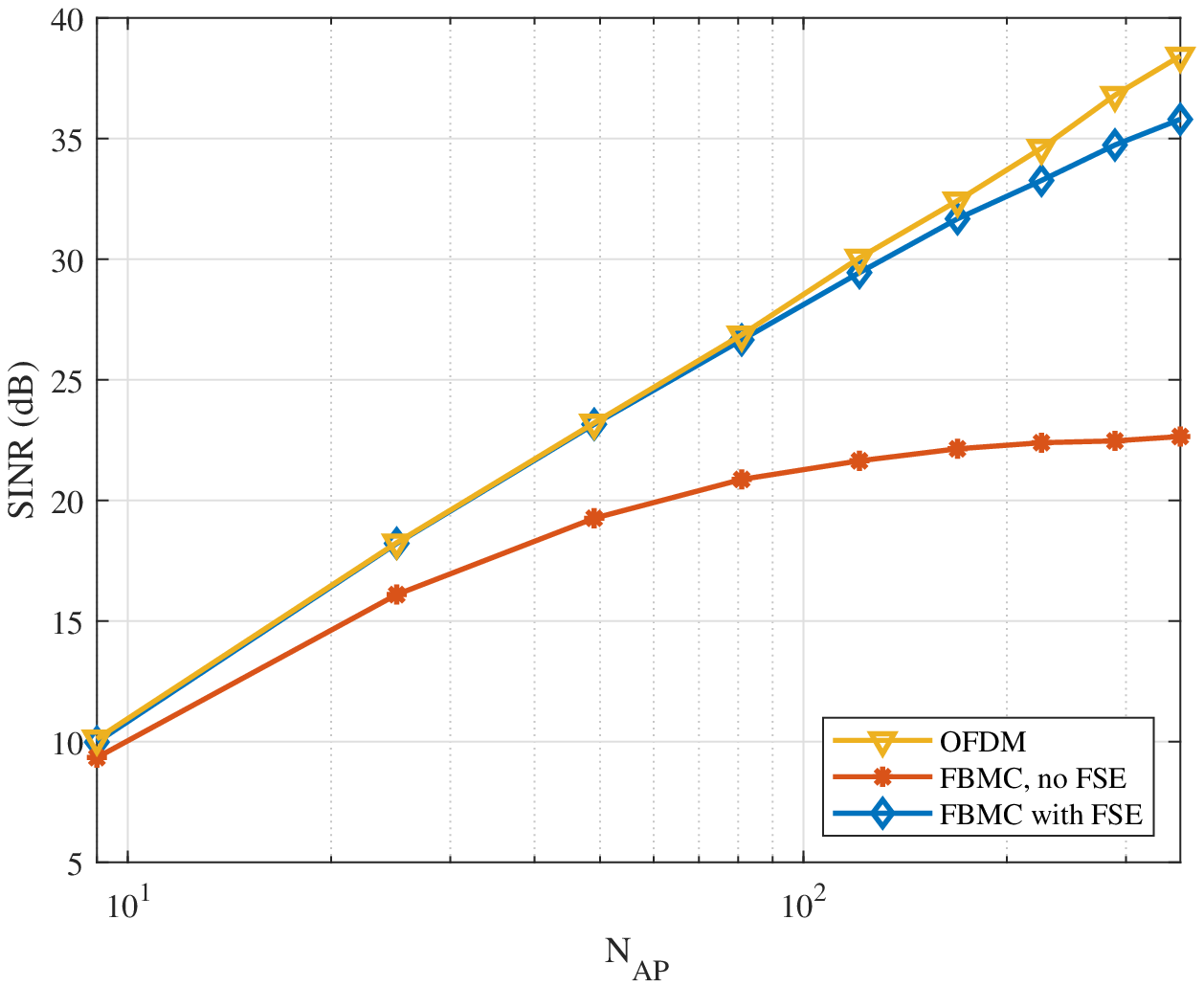}
		\caption{Output SINR vs. number of APs, $N_{AP}$, for FSE design  $L_{\rm FSE}=5$.}
		\label{fig:perfectCSI_cellfree_ZF}
	\end{figure}
	
In Fig.~\ref{fig:perfectCSI_cellfree_ZF}, the SINR of our proposed two-stage FBMC receiver is compared with an OFDM system, as a benchmark. Here, the FSE has a length of $L_{\rm FSE}=5$. CSI is assumed to be known perfectly, and a ZF combiner is used as the first equalization stage at the FBMC receiver. The close performance of FBMC to OFDM confirms its efficacy. The small deviation of FBMC from OFDM, here, is attributed to the short length $L_{\rm FSE}=5$. It can be resolved by increasing $L_{\rm FSE}$ to $9$, as in Fig.~\ref{fig:SINRLength}.

	\begin{figure}
		\centering		\includegraphics[scale=0.61,trim={0 0 0 0},clip]{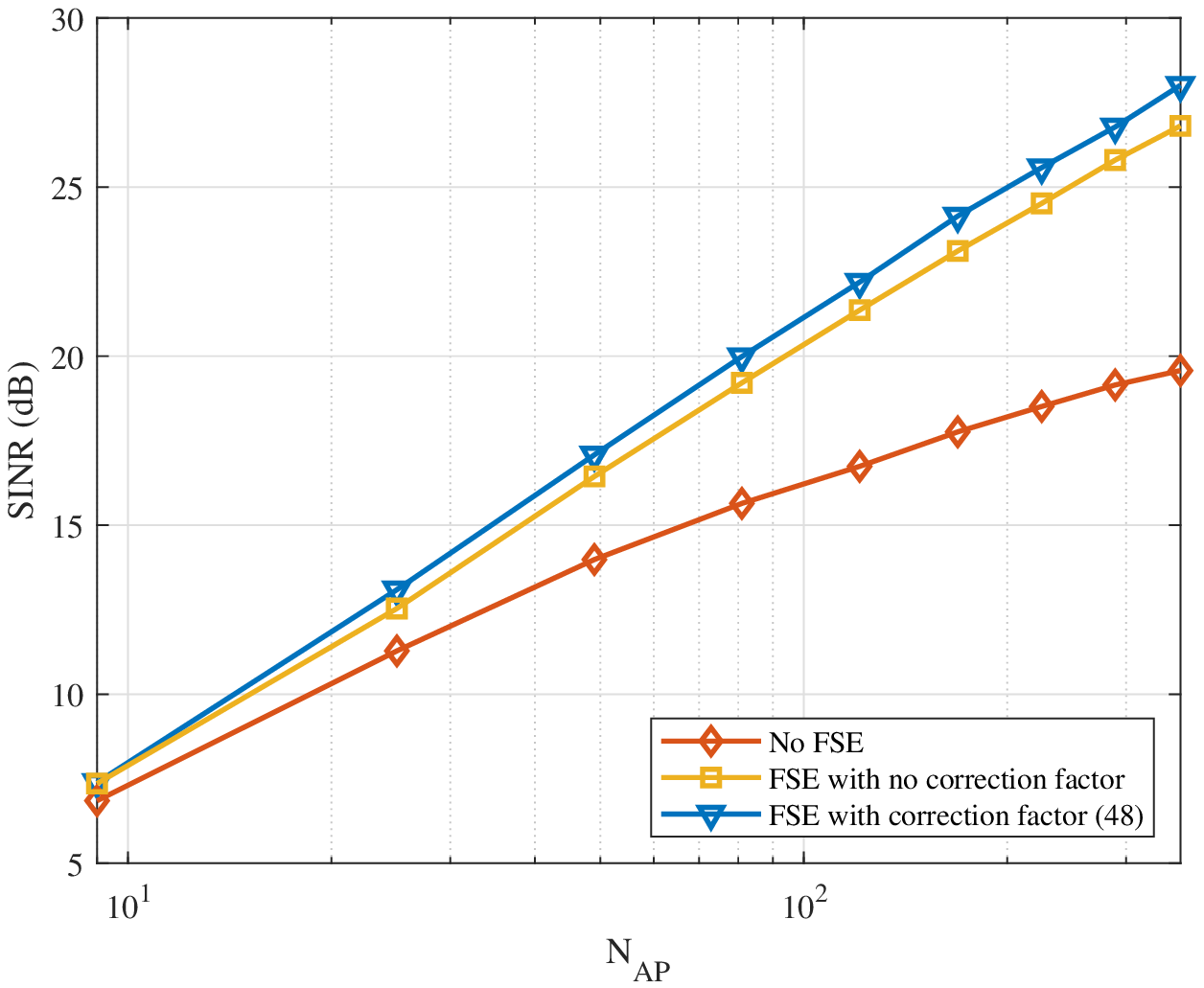}
		\caption{Output SINR vs. number of APs, $N_{AP}$, for FSE design  $L_{\rm FSE}=5$.}
		\label{fig:cellfree_imperfectCSI}
\end{figure}

Fig.~\ref{fig:cellfree_imperfectCSI} presents simulation results for a case where CSI is estimated. The improvement that results from the correction term in \eqref{eq:MMSEinput_est_cell} is also shown. Compared to the case of co-located antennas, the amount of improvement is relatively small. This difference may be attributed to the fact that in a distributed antenna/cell-free scenario, the  number of effective antennas that serve each user remains small, no matter how large the total number of antennas or the number of APs will be. The SINR improvement that is seen in Fig.~\ref{fig:cellfree_imperfectCSI} as $N_{\rm AP}$ increases is due to the fact that for large values of $N_{\rm AP}$, there are always one or more APs near each user. Hence, the input SNR with a larger number of effective antennas can be very large.

\section{Conclusion}
\label{sec:conclusion}
In this work, we designed practical receivers for FBMC-based massive MIMO in both co-located and distributed antenna setups with perfect/imperfect knowledge of CSI. We proposed a spectrally efficient channel estimation method that acquires the channel impulse responses of all the users jointly. Channel estimation error statistics were also calculated and utilized to improve the receiver designs. We proposed a two-stage equalization technique to improve on the performance of the FBMC-based massive MIMO systems. The proposed design consists of a linear combiner followed by a set of FSEs, i.e., one for each subcarrier of each user. We studied three different designs of the FSE. These designs may be thought of as modifications to a previously published work in our group, taking into account a number of needs in practical applications. The emphasis was to (i) reduce the FSE length, hence, save on the computational complexity and minimize the receiver processing latency; (ii) introduce new designs that are applicable to the more general applications, including distributed antenna systems and the recently proposed cell-free network architecture; (iii) take into account the channel estimation errors. Finally, we demonstrated the successful performance of the proposed designs through extensive numerical results.

\appendices

\ifCLASSOPTIONcaptionsoff
  \newpage
\fi

\bibliography{ref.bib}

\begin{thebibliography}{10}
\providecommand{\url}[1]{#1}
\csname url@samestyle\endcsname
\providecommand{\newblock}{\relax}
\providecommand{\bibinfo}[2]{#2}
\providecommand{\BIBentrySTDinterwordspacing}{\spaceskip=0pt\relax}
\providecommand{\BIBentryALTinterwordstretchfactor}{4}
\providecommand{\BIBentryALTinterwordspacing}{\spaceskip=\fontdimen2\font plus
\BIBentryALTinterwordstretchfactor\fontdimen3\font minus
  \fontdimen4\font\relax}
\providecommand{\BIBforeignlanguage}[2]{{%
\expandafter\ifx\csname l@#1\endcsname\relax
\typeout{** WARNING: IEEEtran.bst: No hyphenation pattern has been}%
\typeout{** loaded for the language `#1'. Using the pattern for}%
\typeout{** the default language instead.}%
\else
\language=\csname l@#1\endcsname
\fi
#2}}
\providecommand{\BIBdecl}{\relax}
\BIBdecl

\bibitem{tataria20206g}
H.~Tataria, M.~Shafi, A.~F. Molisch, M.~Dohler, H.~Sj{\"o}land, and
  F.~Tufvesson, ``{6G} wireless systems: Vision, requirements, challenges,
  insights, and opportunities,'' \emph{arXiv preprint arXiv:2008.03213}, 2020.

\bibitem{farhang2011ofdm}
B.~Farhang-Boroujeny, ``{OFDM} versus filter bank multicarrier,'' \emph{IEEE
  signal processing magazine}, vol.~28, no.~3, pp. 92--112, 2011.

\bibitem{nissel2017filter}
R.~Nissel, S.~Schwarz, and M.~Rupp, ``Filter bank multicarrier modulation
  schemes for future mobile communications,'' \emph{IEEE Journal on Selected
  Areas in Communications}, vol.~35, no.~8, pp. 1768--1782, 2017.

\bibitem{farhang2014filter}
B.~Farhang-Boroujeny, ``Filter bank multicarrier modulation: A waveform
  candidate for {5G} and beyond,'' \emph{Advances in Electrical Engineering},
  vol. 2014, 2014.

\bibitem{aminjavaheri2015frequency}
A.~Aminjavaheri, A.~Farhang, N.~Marchetti, L.~E. Doyle, and
  B.~Farhang-Boroujeny, ``Frequency spreading equalization in multicarrier
  massive {MIMO},'' in \emph{2015 IEEE International Conference on
  Communication Workshop (ICCW)}.\hskip 1em plus 0.5em minus 0.4em\relax IEEE,
  2015, pp. 1292--1297.

\bibitem{aminjavaheri2018filter}
A.~Aminjavaheri, A.~Farhang, and B.~Farhang-Boroujeny, ``Filter bank
  multicarrier in massive {MIMO}: Analysis and channel equalization,''
  \emph{IEEE Transactions on Signal Processing}, vol.~66, no.~15, pp.
  3987--4000, 2018.

\bibitem{rottenberg2018performance}
F.~Rottenberg, X.~Mestre, F.~Horlin, and J.~Louveaux, ``Performance analysis of
  linear receivers for uplink massive {MIMO} {FBMC-OQAM} systems,'' \emph{IEEE
  Transactions on Signal Processing}, vol.~66, no.~3, pp. 830--842, 2018.

\bibitem{singh2019uplink}
P.~Singh, H.~B. Mishra, A.~K. Jagannatham, K.~Vasudevan, and L.~Hanzo, ``Uplink
  sum-rate and power scaling laws for multi-user massive {MIMO-FBMC} systems,''
  \emph{IEEE Transactions on Communications}, vol.~68, no.~1, pp. 161--176,
  2019.

\bibitem{schulz2017latency}
P.~Schulz, M.~Matthe, H.~Klessig, M.~Simsek, G.~Fettweis, J.~Ansari, S.~A.
  Ashraf, B.~Almeroth, J.~Voigt, I.~Riedel \emph{et~al.}, ``Latency critical
  {IoT} applications in {5G}: Perspective on the design of radio interface and
  network architecture,'' \emph{IEEE Communications Magazine}, vol.~55, no.~2,
  pp. 70--78, 2017.

\bibitem{lele2008channel}
C.~L{\'e}l{\'e}, J.-P. Javaudin, R.~Legouable, A.~Skrzypczak, and P.~Siohan,
  ``Channel estimation methods for preamble-based {OFDM/OQAM} modulations,''
  \emph{European Transactions on Telecommunications}, vol.~19, no.~7, pp.
  741--750, 2008.

\bibitem{kong2014time}
D.~Kong, D.~Qu, and T.~Jiang, ``Time domain channel estimation for {OQAM-OFDM}
  systems: Algorithms and performance bounds,'' \emph{IEEE Transactions on
  Signal Processing}, vol.~62, no.~2, pp. 322--330, 2014.

\bibitem{caus2012transmitter}
M.~Caus and A.~I. P{\'e}rez-Neira, ``Transmitter-receiver designs for highly
  frequency selective channels in {MIMO FBMC} systems,'' \emph{IEEE
  Transactions on Signal Processing}, vol.~60, no.~12, pp. 6519--6532, 2012.

\bibitem{kofidis2015preamble}
E.~Kofidis, ``Preamble-based estimation of highly frequency selective channels
  in {MIMO-FBMC/OQAM} systems,'' in \emph{Proceedings of European wireless
  2015; 21th European wireless conference}.\hskip 1em plus 0.5em minus
  0.4em\relax VDE, 2015, pp. 1--6.

\bibitem{singh2019time}
P.~Singh and K.~Vasudevan, ``Time domain channel estimation for
  {MIMO-FBMC/OQAM} systems,'' \emph{Wireless Personal Communications}, pp.
  1--20, 2019.

\bibitem{rusek2012scaling}
F.~Rusek, D.~Persson, B.~K. Lau, E.~G. Larsson, T.~L. Marzetta, O.~Edfors, and
  F.~Tufvesson, ``Scaling up {MIMO}: Opportunities and challenges with very
  large arrays,'' \emph{IEEE signal processing magazine}, vol.~30, no.~1, pp.
  40--60, 2012.

\bibitem{ngo2017cell}
H.~Q. Ngo, A.~Ashikhmin, H.~Yang, E.~G. Larsson, and T.~L. Marzetta,
  ``Cell-free massive {MIMO} versus small cells,'' \emph{IEEE Transactions on
  Wireless Communications}, vol.~16, no.~3, pp. 1834--1850, 2017.

\bibitem{hirosaki1980analysis}
B.~Hirosaki, ``An analysis of automatic equalizers for orthogonally multiplexed
  qam systems,'' \emph{IEEE Transactions on Communications}, vol.~28, no.~1,
  pp. 73--83, 1980.

\bibitem{kay1993fundamentals}
S.~M. Kay, \emph{Fundamentals of statistical signal processing}.\hskip 1em plus
  0.5em minus 0.4em\relax Prentice Hall PTR, 1993.

\bibitem{Hoss2006:Spectrally}
H.~Hosseiny, A.~Farhang, and B.~Farhang-Boroujeny, ``Spectrally efficient pilot
  structure and channel estimation for multiuser {FBMC} systems,'' in
  \emph{2020 IEEE International Conference on Communications (ICC): Wireless
  Communications Symposium (IEEE ICC'20 - WC Symposium)}, Dublin, Ireland, Jun.
  2020.

\bibitem{ngo2013energy}
H.~Q. Ngo, E.~G. Larsson, and T.~L. Marzetta, ``Energy and spectral efficiency
  of very large multiuser {MIMO} systems,'' \emph{IEEE Transactions on
  Communications}, vol.~61, no.~4, pp. 1436--1449, 2013.

\bibitem{bjornson2019making}
E.~Bj{\"o}rnson and L.~Sanguinetti, ``Making cell-free massive {MIMO}
  competitive with {MMSE} processing and centralized implementation,''
  \emph{IEEE Transactions on Wireless Communications}, 2019.

\bibitem{1457566}
S.~U.~H. {Qureshi}, ``Adaptive equalization,'' \emph{Proceedings of the IEEE},
  vol.~73, no.~9, pp. 1349--1387, 1985.

\bibitem{farhang2008signal}
B.~Farhang-Boroujeny, \emph{Signal processing techniques for software
  radios}.\hskip 1em plus 0.5em minus 0.4em\relax Lulu publishing house, 2008,
  vol.~2.

\bibitem{vaidyanathan2006multirate}
P.~P. Vaidyanathan, \emph{Multirate systems and filter banks}.\hskip 1em plus
  0.5em minus 0.4em\relax Pearson Education India, 2006.

\bibitem{farhang2013adaptive}
B.~Farhang-Boroujeny, \emph{Adaptive filters: theory and applications}.\hskip
  1em plus 0.5em minus 0.4em\relax John Wiley \& Sons, 2013.

\bibitem{nayebi2015cell}
E.~Nayebi, A.~Ashikhmin, T.~L. Marzetta, and H.~Yang, ``Cell-free massive mimo
  systems,'' in \emph{2015 49th Asilomar Conference on Signals, Systems and
  Computers}.\hskip 1em plus 0.5em minus 0.4em\relax IEEE, 2015, pp. 695--699.

\bibitem{simonsson2008uplink}
A.~Simonsson and A.~Furuskar, ``Uplink power control in {LTE}-overview and
  performance, subtitle: principles and benefits of utilizing rather than
  compensating for {SINR} variations,'' in \emph{2008 IEEE 68th Vehicular
  Technology Conference}.\hskip 1em plus 0.5em minus 0.4em\relax IEEE, 2008,
  pp. 1--5.

\bibitem{nikbakht2020uplink}
R.~Nikbakht, R.~Mosayebi, and A.~Lozano, ``Uplink fractional power control and
  downlink power allocation for cell-free networks,'' \emph{IEEE Wireless
  Communications Letters}, vol.~9, no.~6, pp. 774--777, 2020.

\bibitem{yates1995framework}
R.~D. Yates, ``A framework for uplink power control in cellular radio
  systems,'' \emph{IEEE Journal on selected areas in communications}, vol.~13,
  no.~7, pp. 1341--1347, 1995.

\bibitem{bellanger2010fbmc}
M.~Bellanger, D.~Le~Ruyet, D.~Roviras, M.~Terr{\'e}, J.~Nossek, L.~Baltar,
  Q.~Bai, D.~Waldhauser, M.~Renfors, T.~Ihalainen \emph{et~al.}, ``{FBMC}
  physical layer: a primer,'' \emph{Phydyas}, vol.~25, no.~4, pp. 7--10, 2010.

\bibitem{etsi2017138}
T.~ETSI, ``138 901 v14. 0.0,“5g; study on channel model for frequencies from
  0.5 to 100ghz,'' 3GPP TR 38.901 version 14.0. 0 Release 14),” ETSI, Tech.
  Rep., 2017.

\bibitem{damosso1999cost}
E.~Damosso, L.~M. Correia \emph{et~al.}, ``Cost action 231: Digital mobile
  radio towards future generation systems: Final report,'' \emph{European
  commission}, 1999.

\bibitem{muhammad2009performance}
B.~Muhammad and A.~Mohammed, ``Performance evaluation of uplink closed loop
  power control for lte system,'' in \emph{2009 IEEE 70th Vehicular Technology
  Conference Fall}.\hskip 1em plus 0.5em minus 0.4em\relax IEEE, 2009, pp.
  1--5.

\bibitem{quintero2008advanced}
N.~J. Quintero, ``Advanced power control for utran lte uplink,'' \emph{Master
  of Science Thesis, Aalborg University}, 2008.

\end{thebibliography}
\bibliographystyle{IEEEtran}
%


\end{document}